\documentclass[12pt]{article}

\begin{document}
\newcommand{\nd}[1]{/\hspace{-0.5em} #1}
\begin{titlepage}
\begin{flushright}
{\bf April 2006} \\ 
hep-th/0604175 \\
\end{flushright}
\begin{centering}
\vspace{.2in}
 {\large {\bf Magnon Bound States and the AdS/CFT Correspondence}}

\vspace{.3in}

Nick Dorey\\
\vspace{.1 in}
DAMTP, Centre for Mathematical Sciences \\ 
University of Cambridge, Wilberforce Road \\ 
Cambridge CB3 0WA, UK \\
\vspace{.2in}
%
%
\vspace{.4in}
{\bf Abstract} \\

\end{centering}
We study the spectrum of asymptotic states in the spin-chain
description of planar ${\cal N}=4$ SUSY
Yang-Mills. In addition to elementary magnons, the asymptotic spectrum
includes an infinite tower of multi-magnon bound states with
exact dispersion relation,  
\begin{eqnarray}
\Delta-J_{1} & = &
 \sqrt{Q^{2}+\frac{\lambda}{\pi^{2}}\sin^{2}\left(\frac{p}{2}\right)}, 
\nonumber
\end{eqnarray}
where the positive integer $Q$ is the number of constituent magnons. 
These states account precisely for the known poles in the exact S-matrix. 
Like the elementary magnon, they transform in small representations 
of supersymmetry and are present for all values 
of the 't Hooft coupling. At strong coupling we identify the dual states in
semiclassical string theory.

\end{titlepage}
\paragraph{}
The AdS/CFT correspondence relates the spectrum of free string theory on
$AdS_{5}\times S^{5}$ to the spectrum of operator dimensions in planar 
${\cal N}=4$ SUSY Yang-Mills. Determining this spectrum is an
interesting open problem. Starting from the gauge theory side, the
problem has an elegant reformulation in terms of an integrable spin
chain \cite{MZ, integ} which is diagonalised by the Bethe ansatz (for reviews
see \cite{reviews}). 
Similar integrable structures have also been found in the semiclassical
limit of the dual string theory \cite{BPR} (see also \cite{MSW}). 
In recent work, Hofman and Maldacena (HM) \cite{HM} have emphasised the
importance of a particular limit where the spectrum 
simplifies on both sides of the correspondence (for earlier
work on this limit see \cite{earlier, Janik}).     
In this limit both the 
spin chain and the dual string effectively become very
long. The dynamics can then be analysed in terms of 
asymptotic states and their scattering. 
\paragraph{}
Like the plane-wave
limit \cite{BMN}, the HM limit focuses on operators of large $R$ charge 
$J_{1}$ and scaling dimension $\Delta$ with the difference 
$\Delta-J_{1}$ held fixed. The new feature is that the
$J_{1}$, $\Delta\rightarrow \infty$ limit is taken with the 't Hooft coupling 
$\lambda=g^{2}N$
held fixed. On both sides of the correspondence the HM limit is 
characterised by an 
$SU(2|2)\times SU(2|2)$ supersymmetry with a novel central extension 
\cite{Dynamic}
whose generators act linearly on the worldsheet/spin-chain
excitations. The spectrum consists of elementary
excitations known as magnons which propagate with conserved momentum 
$p$ on the infinite chain. These states are in short representations
of the supersymmetry which essentially determines 
their dispersion relation as\footnote{More precisely supersymmetry 
would allow an arbitrary function of $\lambda$ to multiply the second
term in the square root. However the simple $\lambda$ dependence shown
reproduces all known results both at weak and strong coupling.}   
\cite{BDS, Staud, BS, Dynamic} (see also \cite{Dispersion}), 
\begin{equation}
\Delta-J_{1}=\sqrt{1+\frac{\lambda}{\pi^{2}}\sin^{2}\left(\frac{p}{2}\right)}
\label{d1} 
\end{equation}
The magnon multiplets undergo dispersionless two-body scattering with an 
S-matrix which is also uniquely determined  
by the $SU(2|2)\times SU(2|2)$ supersymmetry up to an overall phase 
\cite{Dynamic}.   
\paragraph{}
In any scattering theory an important possibility is that elementary
excitations can form bound states. Each such object is a new
asymptotic state of the theory with its own dispersion relation and 
S-matrix. Indeed the complete 
spectrum of the theory in the HM limit simply consists of
all possible free multiparticle states including arbitrary numbers of
each species of bound state. 
In this letter we will identify an infinite tower of bound states in 
the $SU(2)$ subsector of the theory. States in this
sector are characterised by a second conserved global charge $J_{2}$
under which the elementary magnon has charge one. Our proposal is that
the full spectrum in the HM limit includes a $Q$-magnon
bound state, for each positive integer $Q$,  
with charge $J_{2}=Q$ and exact dispersion relation, 
\begin{equation}
\Delta-J_{1}=
\sqrt{Q^{2}+\frac{\lambda}{\pi^{2}}\sin^{2}\left(\frac{p}{2}\right)}
\label{d2} 
\end{equation}  
In the context of the full theory we believe that 
these states should give rise to similar 
small representations of supersymmetry as the elementary magnon. In
this context magnon boundstates 
must form complete multiplets the $SO(4)$ subgroup of the R-symmetry  
preserved by the groundstate of the spin chain. The charge $J_{2}$
corresponds to one of the Cartan generators of this group. 
Unlike the bound states discussed in \cite{HM}, they are absolutely 
stable (for non-zero momentum) and exist for all values of the 
't Hooft coupling. The occurrence of an infinite tower of 
BPS bound states is reminiscent of many related phenomena in string
theory and supersymmetric field theory.  
 \paragraph{}
In the rest of the paper we will present evidence for the existence of these
states both in gauge theory and in string theory. For $\lambda<<1$,
our proposal reproduces the well-known spectrum of the 
Heisenberg spin chain in its thermodynamic limit. For $\lambda>>1$ we
will identify bound states of large $Q$ in semiclassical string theory by
taking an appropriate limit of the two-spin folded string solution of 
\cite{FT,AFRT}. However, the most important piece of evidence is valid for
all values of $\lambda$: the $Q=2$ bound state with dispersion relation
(\ref{d2}) accounts {\em precisely} for the known pole in the exact 
two-body S-matrix of \cite{Dynamic, BS}. In the full theory, 
the pole in question has non-trivial matrix structure and is therefore 
associated with the piece of the S-matrix which is uniquely determined
by the supersymmetries. This fits well with the fact that the corresponding
bound states are BPS and their dispersion relation is also uniquely
determined by SUSY. We will also identify the
singularity in the $Q$-body S-matrix corresponding to the $Q$-magnon
bound state. 
\paragraph{}
As in integrable relativistic field theories in
two-dimensions \cite{Zam}, it seems that the bound state spectrum and
its dispersion relation 
places strong constraints on scattering. Further
investigation of the spectrum may be useful in resolving the remaining
ambiguities in the S-matrix. In particular the non-BPS bound states
discussed in \cite{HM} should appear as poles in the as yet
undetermined overall phase of the S-matrix.               
\paragraph{}
We begin by briefly reviewing the spin-chain description of the 
${\cal N}=4$ theory \cite{MZ}. 
The $SU(2)$ sector of ${\cal N}=4$ SUSY Yang-Mills consists of
operators of the form, 
\begin{eqnarray} 
\mathcal{O} & \sim & {\rm Tr}
\left[\Phi_{1}^{J_{1}}\Phi_{2}^{J_{2}}\right] \,\,\,\, +
  \,\,\,\ldots
\nonumber 
\end{eqnarray}
where $\Phi_{1}$ and $\Phi_{2}$ are two of the three complex adjoint
scalars of the theory. The dots denote all possible orderings of the
fields. Each operator in this sector is characterised by two integer
-valued charges $J_{1}$ and $J_{2}$ corresponding to a $U(1)\times
U(1)$ subgroup of the $SU(4)$ R-symmetry group. As usual we will focus
on the planar theory obtained by taking the $N\rightarrow \infty$
limit of the $SU(N)$ theory with the 't Hooft coupling
$\lambda=g^{2}N$ held fixed.    
\paragraph{}
At one-loop, the  
dilatation operator of the theory (in the $SU(2)$ sector) 
can be mapped onto the Hamiltonian
of the Heisenberg spin chain \cite{MZ}. The spectrum
of scaling dimensions can
then be determined by diagonalising the Heisenberg Hamiltonian. More
precisely, at one-loop, the scaling dimension $\Delta$ of an operator
is related to the energy $E$ of the corresponding eigenstate of the
spin-chain as,
\begin{equation}
\Delta=L+\frac{\lambda}{8\pi^{2}}E
\label{one}
\end{equation}
States of the spin chain with charges $J_{1}$ and $J_{2}$ have 
$J_{2}$ flipped spins in a periodic chain of length
$L=J_{1}+J_{2}$. Eigenstates with a single flipped spin are known as
magnons. Magnons have conserved energy $\varepsilon$ and momentum $p$ 
related by the dispersion relation, 
\begin{equation}
\varepsilon(p)=4\sin^{2}\left(\frac{p}{2}\right)
\end{equation}
\paragraph{}
Eigenstates in the sector with $M$ flipped spins are formed as linear 
superpositions of of $M$ magnons. They are characterised by $M$ 
individually conserved momenta $p_{k}$ for $k=1,\ldots, M$ and have total
energy, 
\begin{equation}
E=\sum_{k=1}^{M}\varepsilon(p_{k}) = \sum_{k=1}^{M}\, 
4\sin^{2}\left(\frac{p_{k}}{2}\right)
\end{equation}    
The problem of finding the energy levels is then reduced to
determining the allowed values of the momenta $p_{k}$. These are
determined by the Bethe Ansatz equations, 
\begin{eqnarray}     
\exp\left(iLp_{k}\right)=\, \prod_{j\neq k} \mathcal{S}(p_{k},p_{j}) &
\,\,\,\, &\, \sum_{k=1}^{M} p_{k}=0 
\label{BA}
\end{eqnarray}
for $k=1,\ldots, m$. Here $\mathcal{S}$ is the two-particle S-matrix
which is given as, 
\begin{eqnarray}
\mathcal{S}(p_{k},p_{j}) & = & \frac{\varphi(p_{k})-\varphi(p_{j})+i}
{\varphi(p_{k})-\varphi(p_{j})-i}
\label{smatrix}
\end{eqnarray}
in terms of the phase function $\varphi(p)=\cot(p/2)/2$. 
\paragraph{}
Following \cite{HM}, we now take the limit 
$L\rightarrow \infty$ with $M$ fixed, where we also hold fixed
the momenta $p_{k}$ of individual magnons. This is just the
standard thermodynamic limit of the spin chain 
(see eg \cite{KM,Faddeev}).  As above we will refer to this as the HM limit. 
It is to be contrasted both with the plane-wave or BMN 
limit and the limits appropriate for studying spinning strings where
the momenta $p_{k}$ go to zero with $p_{k}L$ fixed. 
\paragraph{}
The key feature of the HM limit is that, as the 
chain becomes very long, 
the magnons become dilute. 
Thus individual magnons propagate over many sites of the chain between 
interactions and can be thought of as asymptotic states. The
asymptotic states undergo dispersionless two-body scattering with the S-matrix
defined above. In general we may expect that as well as undergoing 
scattering, magnons can form bound states. Roughly
speaking, a $Q$-magnon bound state corresponds to a state of the spin
chain with $Q$ flipped spins where the wavefunction is strongly
peaked on configurations where all the flipped spins are nearly adjacent in
the chain. In the thermodynamic limit where the chain length becomes
infinite this notion becomes more precise and and a bound state can be
defined by demanding a normalisable wavefunction in the usual way. 
Each bound state should then be included as a new asymptotic state of
the scattering theory with its own S-matrix and dispersion law. 
\paragraph{}
For the Heisenberg spin chain, the spectrum of magnon bound states in
the thermodynamic limit is well-known (see Section 5 of
\cite{Faddeev}). 
There is a single bound-state of $Q$ magnons, for each positive
integer $Q\leq L/2$ with dispersion relation, 
\begin{equation}
\varepsilon_{Q}(p)=\frac{4}{Q}\sin^2\left(\frac{p}{2}\right)
\label{disp}
\end{equation}
As we are taking an $L\rightarrow \infty$ limit, this is effectively
an infinite tower. 
\paragraph{}
The recipe for finding these bound states is very simple: two-magnon
bound states correspond to poles in the two-body S-matrix
(\ref{smatrix}) \cite{Faddeev}. In particular, we find such a pole in
$S(p_{1},p_{2})$ when, 
\begin{equation}
\varphi(p_{1})-\varphi(p_{2}) =
\frac{1}{2}\cot\left(\frac{p_{1}}{2}\right)-
\frac{1}{2}\cot\left(\frac{p_{2}}{2}\right) =i
\label{cond}
\end{equation}
which corresponds to a bound state\footnote{This state was also
  discussed briefly in \cite{HM}} with $U(1)$ charge $J_{2}=Q=2$ and
momentum $p=p_{1}+p_{2}$. We solve these conditions by setting
\cite{KM,Faddeev},
\begin{eqnarray} 
p_{1}=\frac{p}{2}+iv & \,\,\,\, & p_{2}=\frac{p}{2}-iv 
\nonumber
\end{eqnarray}
in (\ref{cond}) which yields $\cos(p/2)=\exp(v)$. This yields a state
with energy, 
\begin{equation}
E=\varepsilon(p_{1})+\varepsilon(p_{2})=
4\sin^{2}\left(\frac{p}{4}+i\frac{v}{2}\right)+ 
4\sin^{2}\left(\frac{p}{4}-i\frac{v}{2}\right)=
2\sin^{2}\left(\frac{p}{2}\right)=\varepsilon_{2}(p)
\end{equation}
Thus the position of the pole uniquely fixes the dispersion relation
of the bound state. 
\paragraph{}
The existence of the higher bound states with $Q>2$, and their
dispersion relation (\ref{disp}), can be
inferred from singularities in the multi-particle S-matrix. For any 
integrable spin chain, this is 
given by a product of two-body factors. The corresponding pole
appears when the momenta of the $Q$ constituent magnons satisfy
\cite{Faddeev, BT}, 
\begin{equation}
\varphi(p_{j})-\varphi(p_{j+1})=i
\label{polecondQ}
\end{equation}
for $j=1,2,\ldots Q-1$. This condition is easily solved and leads
directly to the bound-state dispersion relation (\ref{disp}).  
\paragraph{}
So far we have only discussed the spectrum of the theory at one-loop
and only in the $SU(2)$ sector. However, assuming integrability and
the spin chain description persists in the full quantum theory, 
supersymmetry yields powerful constraints on the the magnon dispersion
relation and two-body S-matrix \cite{Dynamic}. 
These constraints provide confirmation for an earlier proposal
\cite{AFS} for an exact Bethe ansatz in the $SU(2)$ sector. As before the 
energy of an M-magnon state is the sum of the energies of individual 
magnons. However, the new ansatz incorporates the exact magnon dispersion
relation,  
\begin{equation}
\varepsilon(p)=\frac{8\pi^{2}}{\lambda}
\left[\sqrt{1+\frac{\lambda}{\pi^{2}}\sin^{2}\left(\frac{p}{2}\right)}-
1\right]
\label{disp1}
\end{equation}
which, because of (\ref{one}), is equivalent to (\ref{d1}).   
The two-body S-matrix which enters in the Bethe ansatz equations 
(\ref{BA}) now takes the form, 
\begin{eqnarray}
\mathcal{S}(p_{k},p_{j}) & = & \frac{\varphi(p_{k})-\varphi(p_{j})+i}
{\varphi(p_{k})-\varphi(p_{j})-i}\times \mathcal{S}_{D}(p_{k},p_{j})
\label{smatrix2}
\end{eqnarray}
where phase function $\varphi(p)$ is now corrected to, 
\begin{equation}
\varphi(p)=\frac{1}{2}\cot\left(\frac{p}{2}\right)
\sqrt{1+\frac{\lambda}{\pi^{2}}\sin^{2}\left(\frac{p}{2}\right)}
\end{equation}
The first factor in (\ref{smatrix2}) originates in the all-loop gauge
theory ansatz of \cite{BDS}. It also appears in the SU(2)
subsector of the full
$SU(2|2)\times SU(2|2)$ S-matrix\footnote{In fact the S-matrix 
factors corresponding to the two $SU(2|2)$ subgroups 
each have a single pole leading to a
double pole in their product (see eqn (40) of \cite{Janik}). 
Thus, the undetermined overall phase
factor should have a simple zero to obtain the expected simple pole in
the complete S-matrix. The author thanks Juan Maldacena for clarifying
this point.} of \cite{Dynamic, Staud, BS}.  
In contrast $\mathcal{S}_{D}$ is a ``dressing factor'' which is related to the
undetermined overall phase of the full S-matrix. 
A formula for $\mathcal{S}_{D}$ was conjectured in \cite{AFS}
which passes many non-trivial tests but we will not need this here. 
The only fact we will use is that the dressing factor 
does not cancel the S-matrix pole which appears in (\ref{smatrix2}) 
when, $\varphi(p_{k})-\varphi(p_{j})=i$.  
\paragraph{}
An obvious question is what happens to the $Q$-magnon bound states
and their dispersion law (\ref{disp}) described above when we move
away from weak coupling. Our proposal is that they
survive for all values of the coupling and have the 
exact dispersion relation,   
\begin{equation}
\varepsilon_{Q}(p)=\frac{8\pi^{2}}{\lambda}
\left[\sqrt{Q^2+\frac{\lambda}{\pi^{2}}\sin^{2}\left(\frac{p}{2}\right)}-
Q\right] 
\label{disp2}
\end{equation}  
which is equivalent to (\ref{d2}). 
This formula clearly reduces to the dispersion relation (\ref{disp}) of the
Heisenberg spin chain at weak coupling. Setting $Q=1$ we obtain the
exact magnon dispersion relation (\ref{disp1}). For $Q=2$, the
proposed bound state should correspond to the pole in the exact
two-body S-matrix (\ref{smatrix2}). Indeed, as above, the pole
position should determine the dispersion relation exactly. We will now
verify this explicitly.   
\paragraph{}
For magnon momenta $p_{1}$ and
$p_{2}$ the new pole condition reads,  
\begin{equation}
\frac{1}{2}\cot\left(\frac{p_{1}}{2}\right)
\sqrt{1+\frac{\lambda}{\pi^{2}}\sin^{2}\left(\frac{p_{1}}{2}\right)}
-\frac{1}{2}\cot\left(\frac{p_{2}}{2}\right)
\sqrt{1+\frac{\lambda}{\pi^{2}}\sin^{2}\left(\frac{p_{2}}{2}\right)}=i
\label{polecond}
\end{equation}
as before we set, 
\begin{eqnarray} 
p_{1}=\frac{p}{2}+iv & \,\,\,\, & p_{2}=\frac{p}{2}-iv 
\nonumber
\end{eqnarray}
and solve for the bound state momentum $p=p_{1}+p_{2}$ as a function of
$v$. After some computation we obtain a sixth-order polynomial
equation, $P_{6}(t)=0$, in $t=\cos(p/2)$ with coefficients polynomial in 
$\exp(v)$ and $a=\lambda/4\pi^{2}$. The polynomial $P_{6}(t)$ can be factored
exactly into the product of a quadratic $P_{2}(t)$ and a quartic
$P_{4}(t)$ which are conveniently given as,  
\begin{eqnarray}
P_{2}(t) & = &  a(e^{2v}-1)^{2}(1+e^{2v}-2e^{v}t)^{2} 
-4e^{2v}(1+6e^{2v}+e^{4v}-4e^{v}t-4e^{3v}t) \nonumber 
\\
P_{4}(t) &= & a(1+e^{2v}-2e^{v}t)^{2}(t^{2}-1)+4e^{v}\left(
t+e^{2v}t-e^{v}(1+t^{2})\right)
\label{quartic}
\end{eqnarray}
The physical root is singled out by its weak-coupling behaviour
$t=\exp(v)$ needed for agreement with the corresponding formula for the 
Heisenberg spin chain 
discussed above. Taking the limit $a\rightarrow 0$, 
one may easily check that the physical root 
belongs to the quartic equation $P_{4}(t)=0$ rather than the quadratic.   
\paragraph{}
The next step is to extract the physical root of the quartic
$P_{4}(t)=0$, use it to eliminate $v$ in the energy formula, 
\begin{eqnarray}
\varepsilon_{2}(p) &= & \varepsilon(p_{1})+\varepsilon(p_{2}) \nonumber \\
      & = & \frac{8\pi^{2}}{\lambda}
\left[\sqrt{1+\frac{\lambda}{\pi^{2}}\sin^{2}
\left(\frac{p}{4}+i\frac{v}{2}\right)}+\sqrt{1+\frac{\lambda}{\pi^{2}}\sin^{2}
\left(\frac{p}{4}-i\frac{v}{2}\right)}-
2\right] \nonumber 
\end{eqnarray}
and compare with the predicted dispersion relation (\ref{disp2})
for the $Q=2$ case. A necessary and sufficient condition for 
agreement with (\ref{disp2}) is that the physical root 
of the quartic should also obey the corresponding energy
conservation equation, 
\begin{equation}
\sqrt{1+\frac{\lambda}{\pi^{2}}\sin^{2}
\left(\frac{p}{4}+i\frac{v}{2}\right)}+\sqrt{1+\frac{\lambda}{\pi^{2}}\sin^{2}
\left(\frac{p}{4}-i\frac{v}{2}\right)}= \sqrt{4+\frac{\lambda}{\pi^{2}}\sin^{2}
\left(\frac{p}{2}\right)}
\end{equation}
Squaring this equation twice and rewriting it in terms of
$t=\cos(p/2)$, $\exp(v)$ and $a=\lambda/4\pi^{2}$ we obtain the
same quartic equation $P_{4}(t)=0$, with $P_{4}$ as in
(\ref{quartic}) and we are done. 
As for the Heisenberg spin chain, the multi-particle S-matrix has a
pole corresponding to a $Q$-magnon bound state for each $Q$ when the
condition (\ref{polecondQ}) is satisfied. In principle 
we could check our proposed dispersion 
relation (\ref{disp2}) for $Q>2$ by solving this condition, 
but we will not pursue this here.  
\paragraph{}
The magnon bound states described above correspond to string theory
states\footnote{More precisely, to obtain an allowed state of the 
closed string we 
should consider two or more excitations with total momenta equal to
zero. As mentioned in \cite{HM}, the central charge vanishes on such 
multiparticle states and so they cannot be BPS. However the exact
energies of these states in the HM limit are simply the sum of the
energies of their constituent magnons which are almost free in this
limit.} with energy, momentum and angular momenta related as, 
\begin{equation}
\Delta-J_{1}=
\sqrt{J_{2}^{2}+\frac{\lambda}{\pi^{2}}\sin^{2}\left(\frac{p}{2}\right)}
\label{d3} 
\end{equation}
For $p=0$, the bound state saturates the
familiar BPS bound, $\Delta\geq J_{1}+J_{2}$, of the full 
$SU(2,2|4)$ superalgebra. For non-zero momentum the state
appears to lie above the bound. However, as explained in \cite{HM},
this is not the case because the magnon momentum can appear as a 
central extension of the supersymmetry algebra which modifies the BPS
bound. Indeed the formula (\ref{d3}) appears to be precisely the
relevant BPS condition for all values of $J_{2}$, generalising the
single magnon result of \cite{HM}. For this reason it seems likely
that all the magnon bound states discussed above give rise to similar small
representations of supersymmetry as the elementary magnon.    
\paragraph{}
To test the proposed spectrum of bound states further we will now go to
the regime of fixed large 't Hooft coupling, $\lambda>>1$ and look for
the corresponding states in semiclassical string theory. 
The HM limit is one where the energy, $\Delta$, of the string
state and one of its angular momenta $J_{1}$ both go to infinity with 
the difference $\Delta-J_{1}$ (and $\lambda$) held fixed. As in the
spin chain, this thermodynamic limit is taken holding the 
momenta and other quantum numbers of individual world-sheet
excitations fixed. 
\paragraph{}
In \cite{HM} the above limit was taken for strings moving on an
$R\times S^{2}$ subspace of $AdS_{5}\times S^{5}$ carrying a single
non-zero angular momentum $J_{1}$. Classical solutions were presented
corresponding to magnons of arbitrary momentum. In general the
solutions correspond to folded strings with endpoints on the equator of
$S^2$. One particularly simple case is that of momentum
$p=\pm\pi$ where the dispersion relation (\ref{d1})
corresponds to a
stationary particle on the string. A consistent state in closed string
theory can be built by taking two such magnons with momenta
$p=\pm\pi$. This corresponds to a special case 
of the folded spinning string solution of Gubser, Klebanov and
Polyakov (GKP) \cite{GKP}. In this limit,
the string rotates around the north pole on $S^{2}$ with its endpoints
moving around the equator at the speed of light. The classical energy
of this state is infinite as is its angular momentum, but the
difference $\Delta-J_{1}$ is finite and equal to
$2\sqrt{\lambda}/\pi$. This matches the expected energy of the 
two magnon configuration described above.       
\paragraph{}
In the following we will study a simple generalisation of this case
with two non-zero angular momenta $J_{1}$ and $J_{2}$. Our starting
point is the two-spin generalisation of the GKP solution 
first presented in \cite{FT}. This corresponds to a string moving on 
an $R\times S^{3}$ subspace of $AdS_{5}\times S^{5}$. 
String motion is described by a four-component vector 
$\vec{X}(\sigma,\tau)=(X_{1},X_{2},X_{3},X_{4})$ of unit length, 
$|\vec{X}|^{2}=1$ which specifies a point on $S^{3}\subset S^{5}$. 
The additional time coordinate is eliminated with the static gauge
condition $X_{0}=\kappa\tau$. The relevant configuration, which
corresponds to a
genus-two finite gap solution of the $SU(2)$ principal chiral model,
can be found using the the ansatz, 
\begin{eqnarray}
X_{1}+iX_{2}=x_{1}(\sigma)\exp(i\omega_{1}\tau)
& \quad{} &  X_{3}+iX_{4}=x_{2}(\sigma)\exp(i\omega_{2}\tau)  
\nonumber
\end{eqnarray}
with $x_{1}(\sigma)^{2}+x_{2}(\sigma)^{2}=1$. The string has energy
$\Delta=\sqrt{\lambda}\kappa$ and conserved angular
momenta, 
\begin{eqnarray}
J_{1}=\sqrt{\lambda}\omega_{1}\int_{0}^{2\pi}\, \frac{d\sigma}{2\pi}\,
x_{1}(\sigma)^{2} & \,\,\, &
J_{2}=\sqrt{\lambda}\omega_{2}\int_{0}^{2\pi}\, 
\frac{d\sigma}{2\pi}\,x_{2}(\sigma)^{2}
\end{eqnarray}
 \paragraph{}
The solution corresponding to a folded spinning string is \cite{FT, AFRT}, 
\begin{eqnarray}
x_{1}=k\, {\rm sn}\left(A\sigma, k\right) & \qquad{} & x_{2}={\rm dn}
\left(A\sigma, k\right)
\end{eqnarray}
with, 
\begin{eqnarray}
A=\sqrt{\omega_{1}^{2}-\omega_{2}^{2}} & \qquad{} & k=
\sqrt{\frac{\kappa^{2}-\omega^{2}_{2}}{\omega_{1}^{2}-\omega_{2}^{2}}}\,\leq
1
\end{eqnarray}
The Jacobian elliptic functions ${\rm sn}$ and ${\rm dn}$ are defined
according to the conventions of \cite{GR}. The closed string
boundary condition $\sigma\sim \sigma+2\pi$ yields the relation 
\begin{equation}
 A=\frac{2}{\pi}K\left(k\right)
\end{equation}
Evaluating the angular momenta on this
solution we obtain, 
\begin{eqnarray}
J_{1}=\sqrt{\lambda}\omega_{1}
\left[1-
\frac{E(k)}{K(k)}\right]
 & \,\,\, &
J_{2}=\sqrt{\lambda}\omega_{2}\frac{E(k)}{K(k)}
\end{eqnarray}
where $K$ and $E$ are complete elliptic integrals. 
\paragraph{}
It is convenient to introduce the variable $\rho=\omega_{2}/\omega_{1}<1$. 
\begin{eqnarray}
\Delta & = & \frac{2\sqrt{\lambda}}{\pi}\, \frac{\sqrt{\rho^{2}+k^{2}
(1-\rho^{2})}}{
\sqrt{1-\rho^{2}}}\, K(k) \nonumber \\ 
J_{1} & = & \frac{2\sqrt{\lambda}}{\pi}\,
\frac{1}{\sqrt{1-\rho^{2}}}\, \
\left(K(k)-E(k)\right) \nonumber \\ 
J_{2}& = & \frac{2\sqrt{\lambda}}{\pi}\,
\frac{\rho}{\sqrt{1-\rho^{2}}}\, E(k)
\end{eqnarray}
\paragraph{}
We will now consider a limit of the HM type where $\Delta\rightarrow \infty$
and $J_{1}\rightarrow \infty$ with the difference $\Delta-J_{1}$ held
fixed. We will also hold the parameter $\rho$ fixed. 
To this end we take take $k\rightarrow 1$ so that  
\begin{eqnarray}  
K(k) & \simeq & -\frac{1}{2}\, \log(1-k) \,\,\rightarrow \infty 
\nonumber 
\end{eqnarray}
and $E(k)\rightarrow 1$. In the limit we find the formulae, 
\begin{eqnarray}   
\Delta-J_{1} = \frac{2\sqrt{\lambda}}{\pi}\frac{1}{\sqrt{1-\rho^{2}}}  
& \,\,\,\, & 
J_{2}= \frac{2\sqrt{\lambda}}{\pi}\frac{\rho}{\sqrt{1-\rho^{2}}}
\end{eqnarray}
This corresponds a one-parameter generalisation of the limiting GKP
solution considered in \cite{HM}, the latter being the special case $\rho=0$. 
Eliminating  the remaining parameter $\rho$, we obtain the relation, 
\begin{equation}
\Delta-J_{1}=
2\sqrt{\left(\frac{J_{2}}{2}\right)^{2}+\frac{\lambda}{\pi^{2}}}
\label{result}
\end{equation}
As the folded string configuration is symmetric it is natural to
interpret this state as consisting of two excitations each carrying
half the total transverse angular momentum $J_{2}$. As before the two
states have momenta $p=\pm\pi$. If we identify each of these excitations as
bound states of $J_{2}/2$ magnons, where $J_{2}\sim \sqrt{\lambda}$ 
then the result (\ref{result}) agrees
with the expected total energy calculated using the proposed
dispersion relation (\ref{d2}).   
\paragraph{}
I would like to thank Gleb Arutyunov, Niklas Beisert and Sergey Frolov 
for helpful comments. 
The author is supported by a PPARC Senior Fellowship.

\end{document}